\documentstyle[11pt,newpasp,twoside,epsf]{article}
\def\edcomment#1{\iffalse\marginpar{\raggedright\sl#1\/}\else\relax\fi}
\reversemarginpar

\begin{document}
\title{Deconfinement signals from pulsar timing}
\author{Hovik Grigorian, Gevorg Poghosyan and Edvard Chubarian}
\affil{Physics Department, Yerevan State University, Alex
        Manoogian Str. 1, 375025 Yerevan, Armenia}
\author{David Blaschke}
\affil{Fachbereich Physik, Universit\"at Rostock,
        Universit\"atsplatz 1, D--18051 Rostock, Germany}

\begin{abstract}
The occurence of a quark matter core in rotating compact stars  
has been investigated within general relativity
as a function of both the rotational frequency and the total baryon number.
%The contributions to the moment of inertia have been evaluated separately. 
%with the result that the changes due to shape deformation and 
%matter redistribution are the dominant ones.
We demonstrate that the deviation of the braking index from $n=3$ signals not 
only the occurence but also the size of a quark matter core in a pulsar.
We suggest that in systems with mass accretion onto a rapidly 
rotating compact star a spin-down to spin-up transition might signal a 
deconfinement transition in its interior. 
\end{abstract}

%\section{Introduction}
Recently it has been suggested that the deconfinement transition in rapidly 
rotating compact stars could have observable consequences in pulsar timing 
since phase transitions in the dense stellar interior could result in changes 
of the rotational behavior (Glendenning, Pei, \& Weber 1997;
Grigorian, Hermann, \& Weber 1999). 
Further constraints for the nuclear equation of state come from the 
observation of quasi-periodic brightness oscillations (QPO's) in low-mass 
X-ray binaries which entail mass and radius limits for rapidly rotating 
neutron stars (Lamb, Miller, \& Psaltis 1998).   

We apply the method of perturbation theory (Hartle 1967; Hartle \& Thorne 1968;
Sedrakian \& Chubarian 1968) which is the most transparent and systematic 
approach to the problem of stationary gravitational fields and their sources 
and has been proven successful in general relativity as well as in alternative 
theories of gravitation (Grigorian \& Chubarian 1985).

We have calculated the mass, angular momentum and shape deformation
from the iterative solution of the gravitational field equations in
case of hydrodynamical, thermodynamical and chemical
equilibrium for given total baryon number and angular velocity
$\Omega$ of the object with a model equation of state describing the 
deconfinement transition (Chubarian et al. 1999).

The moment of inertia changes predominantly due to matter redistribution and 
shape deformation rather than gravitational fields and rotational energy and 
reflects the occurrence of a quark matter core (Chubarian et al. 1999). 
Observable consequences of the onset of deconfinement are changes in the 
braking index for isolated pulsars, see Fig. 1 (a), 
and a flip from spin-down to spin-up behavior, see Fig. 1 (b), for
systems with mass accretion at constant angular momentum $J$.   

\begin{figure}[ht]
\plottwo{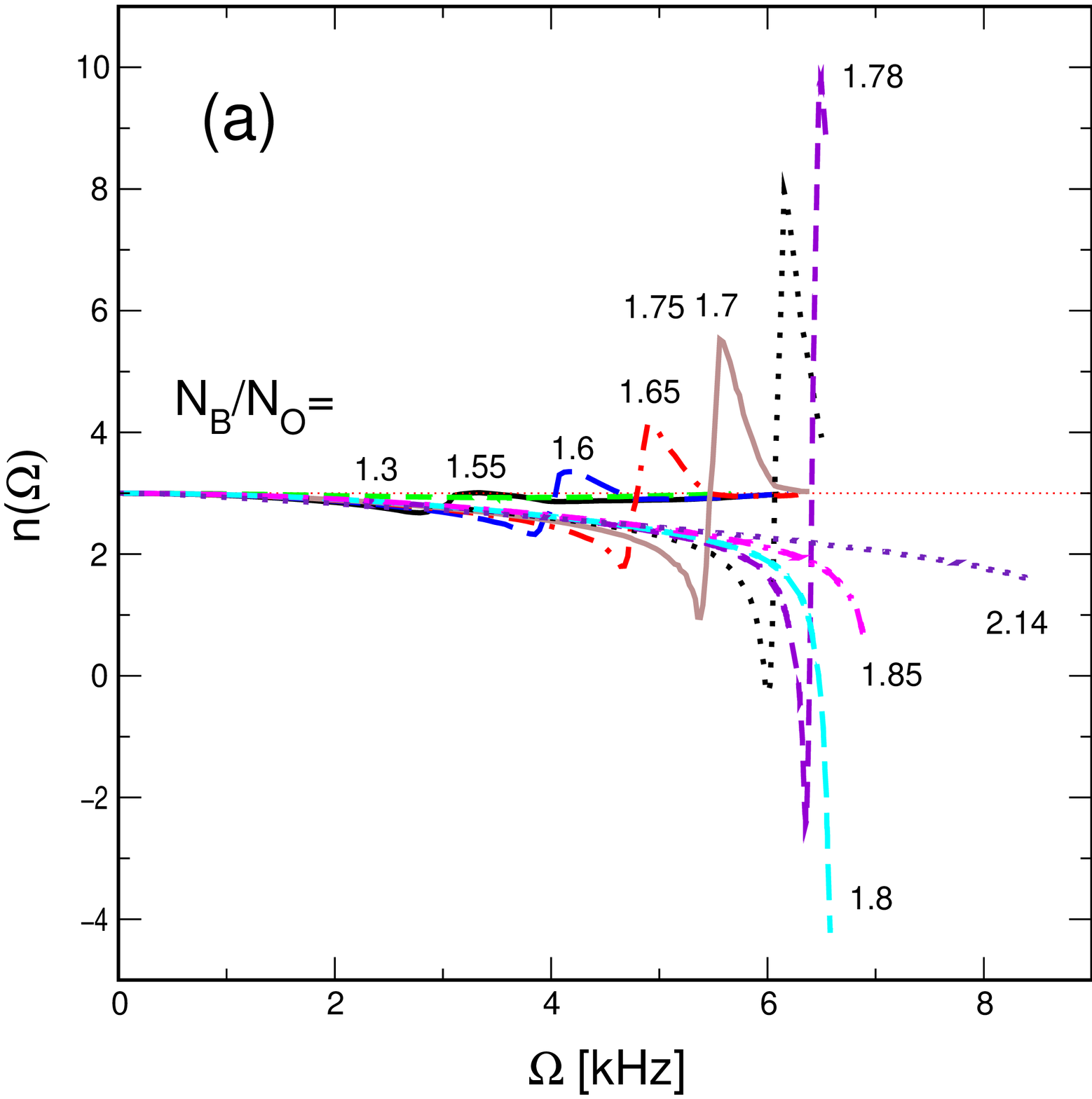}{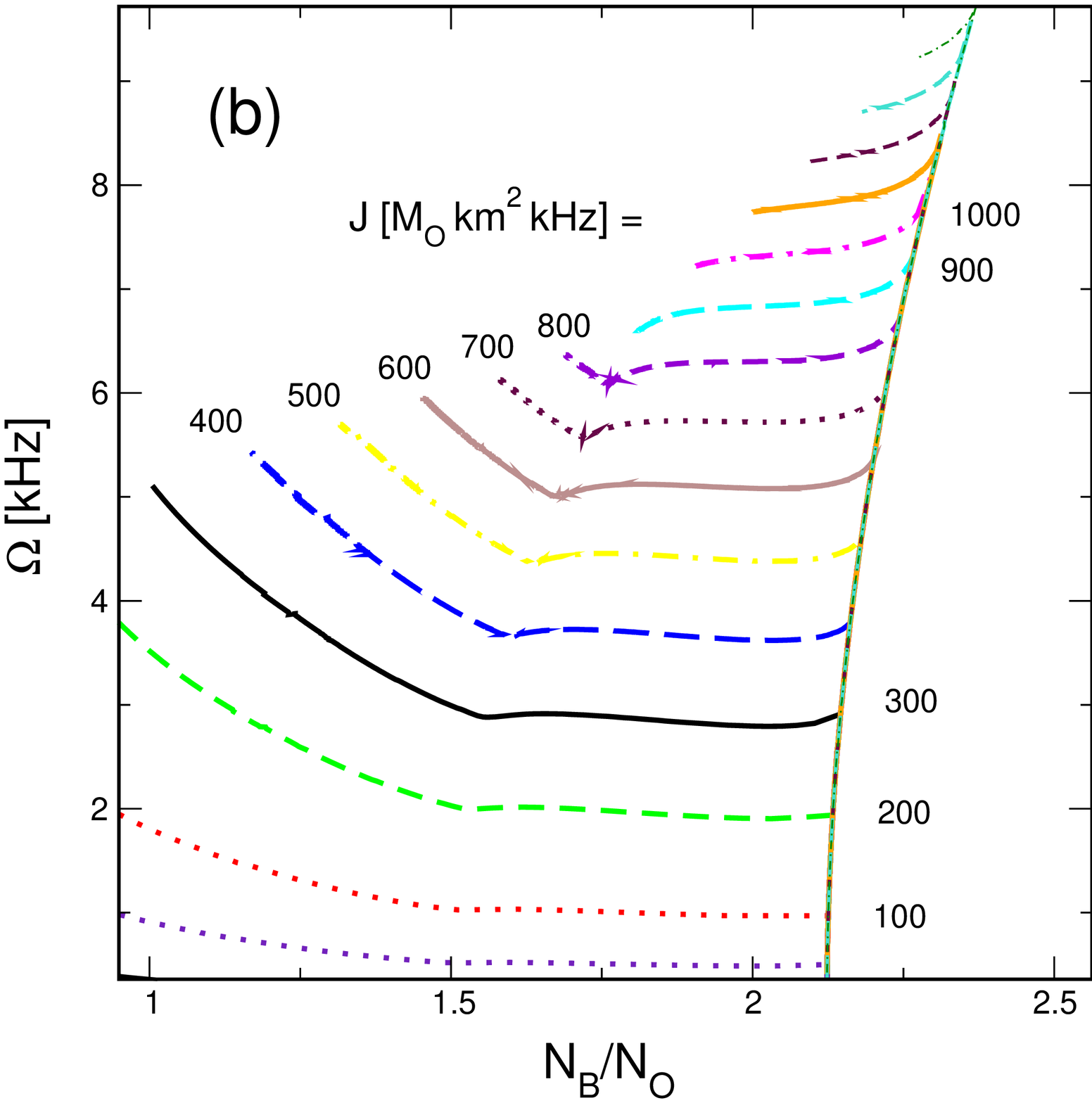}
\caption{(a) Braking index $n(\Omega)$ for configurations with different 
total baryon numbers. Deviation from $n=3$ indicates the occurrence of a quark 
matter core correlated with its size. (b) Spin-down to spin-up transition 
for mass accretion at constant angular momentum $J$ as a function of the 
baryon number $N_B$ accumulated in the compact star.
\label{timing}}
\end{figure}

Comparison of our results (Fig. 1 (a)) with those of Glendenning, Pei \& Weber 
(1997) shows that deconfinement signals from pulsar timing are very sensitive 
to the choice of the EOS and we have therefore started QCD based studies of the
EOS at finite baryon density and temperature for applications to compact stars
(Blaschke et al. 1999). 
  
{\acknowledgements}
This work has been supported by the Volkswagen Stiftung under
grant no.\ I/71 226 and by a stipend from the DAAD.

\end{document}